\documentclass [prl,twocolumn,longbibliography]{revtex4-1}

\usepackage{graphicx}
\usepackage{subfigure}
\usepackage{amsmath}
\usepackage{amsthm}
\usepackage{amssymb}
\usepackage{hyperref}
\usepackage{xcolor}
\usepackage{textcomp}
\usepackage[normalem]{ulem}

\hypersetup{colorlinks=true, linkcolor=black, citecolor=black, urlcolor=blue}

\newcommand{\ket}[1]{\left| #1 \right>} % for Dirac bras
\newcommand{\bra}[1]{\left< #1 \right|} % for Dirac kets
\begin{document}

%=========================================================================
\title{Conventional and unconventional Dicke models: Multistabilities and nonequilibrium dynamics}
 
\author{Farokh Mivehvar}
\email[Corresponding author: ]{farokh.mivehvar@uibk.ac.at}
\affiliation{Institut f\"ur Theoretische Physik, Universit{\"a}t Innsbruck, A-6020~Innsbruck, Austria}

\begin{abstract}
The Dicke model describes the collective behavior of a sub-wavelength--size ensemble of two-level atoms (i.e., spin-1/2) interacting identically with a single quantized radiation field of a cavity. Across a critical coupling strength it exhibits a zero-temperature phase transition from the normal state to the superradiant phase where the field is populated and the collective spin acquires a nonzero $x$-component, which can be imagined as \emph{ferromagnetic} ordering of the atomic spins along $x$. Here we introduce a variant of this model where two sub-wavelength--size ensembles of spins interact with a single quantized radiation field with different strengths. Subsequently, we restrict ourselves to a special case where the coupling strengths are opposite (which is unitarily equivalent to equal-coupling strengths). Due to the conservation of the total spin in each ensemble individually, the system supports two distinct superradiant states with $x$-ferromagnetic and $x$-\emph{ferrimagnetic} spin ordering, coexisting with each other in a large parameter regime. The stability and dynamics of the system in the thermodynamic limit are examined using a semiclassical approach, which predicts non-stationary behaviors due to the multistabilities. At the end, we also perform small-scale full quantum-mechanical calculations, with results consistent with the semiclassical ones. 
\end{abstract}

\maketitle

%=========================================================================
\emph{Introduction.}---The Dicke model is one of the most celebrated models in quantum optics~\cite{Hepp1973On, Hepp1973Equilibrium, Wang1973}. It describes the collective behavior of $N$ two-level atoms (i.e., spin-1/2) cooperatively interacting with a single quantized radiation field~\cite{Larson2021}. The model has a fairly simple Hamiltonian (we set $\hbar=1$ throughout the paper)~\cite{Garraway2011},
\begin{align} \label{eq:Dicke-H}
\hat{H}_{\rm D}=\omega_c\hat{a}^\dag\hat{a}+\omega_a\hat{S}_z+\lambda(\hat{a}^\dag+\hat{a})\hat{S}_x,
\end{align}
where $\omega_c$ is the frequency of the cavity mode, $\omega_a$ the atomic transition frequency, and $\lambda$ the single-atom--field coupling strength. Here, $\hat{a}$ is the bosonic annihilation operator of the cavity radiation field and $\hat{\mathbf S}=(\hat{S}_x,\hat{S}_y,\hat{S}_z)$ is the collective atomic spin operator of maximum length $N/2$ with components $\hat{S}_\alpha=\sum_{j=1}^N \hat{s}_{j,\alpha}$, where $\alpha=\{x,y,z\}$, defined in terms of the common single-atom spin operators $\hat{s}_{j,\alpha}$. Despite its simple form, the Dicke Hamiltonian \eqref{eq:Dicke-H} is expected to exhibit a variety of interesting phenomena, most notably the zero-temperature phase transition from the normal (N) state with the field being in the vacuum state to the superradiant (SR) phase where the field acquires a nonzero photon number $\langle \hat{a}^\dag\hat{a} \rangle \neq 0$ across the critical coupling strength $\sqrt{N}\lambda_c=\sqrt{\omega_c\omega_a}$~\cite{Kirton2018}. Correspondingly, the collective atomic spin completely in the spin-down state $\langle \hat{S}_z \rangle = -N/2$ initially---i.e., \emph{ferromagnetically} (Fo) ordered along the negative $z$ direction, referred to as ``$-$zFo-N'' in this work---obtains a nonzero $x$-component  $\langle \hat{S}_x \rangle \neq 0$. The collective atomic spin orients almost completely in the positive or negative $x$ direction $\langle \hat{S}_x \rangle \to \pm N/2$, depending on the broken $\mathbb{Z}_2$ parity symmetry of the Dicke Hamiltonian on the onset of the superradiant phase transition, in the deep superradiant phase $\lambda\to\infty$. Therefore, we refer to this superradiant phase with its asymptotic ferromagnetic ordering along the positive or negative $x$ direction as ``$+$xFo-SR" and ``$-$xFo-SR", respectively, designated collectively by ``$\pm$xFo-SR" (the ``$+$" and ``$-$'' signs specify the direction of the total spin along a desired axis).

As proposed theoretically~\cite{Domokos2002, Dimer2007, Nagy2010, Safaei2013, Joshi2015, Mivehvar2017, Mivehvar2019, Chiacchio2019, Buca2019, Masalaeva2021, Chiacchio2023, Masalaeva2023}, the Dicke model has been successfully implemented using cavity-assisted two-photon Raman transitions between low-lying atomic momentum and/or hyperfine states for both bosonic~\cite{Baumann2010, Klinder2015, Kollr2017, Zhiqiang2017, Landini2018, Kroeze2018, Naik2018, Kroeze2019, Dogra2019} and fermionic~\cite{Zhang2021, Helson2023} atoms, bypassing the no-go theorems~\cite{Bialynicki-Birula1979, Nataf2010}. There is also great interest in realizing Dicke-type models and superradiance in waveguide-QED setups~\cite{Chang2018,Sheremet2023} and cavity quantum materials~\cite{Schlawin2022}. 

\begin{figure}[t!]
\centering
\includegraphics [width=0.48\textwidth]{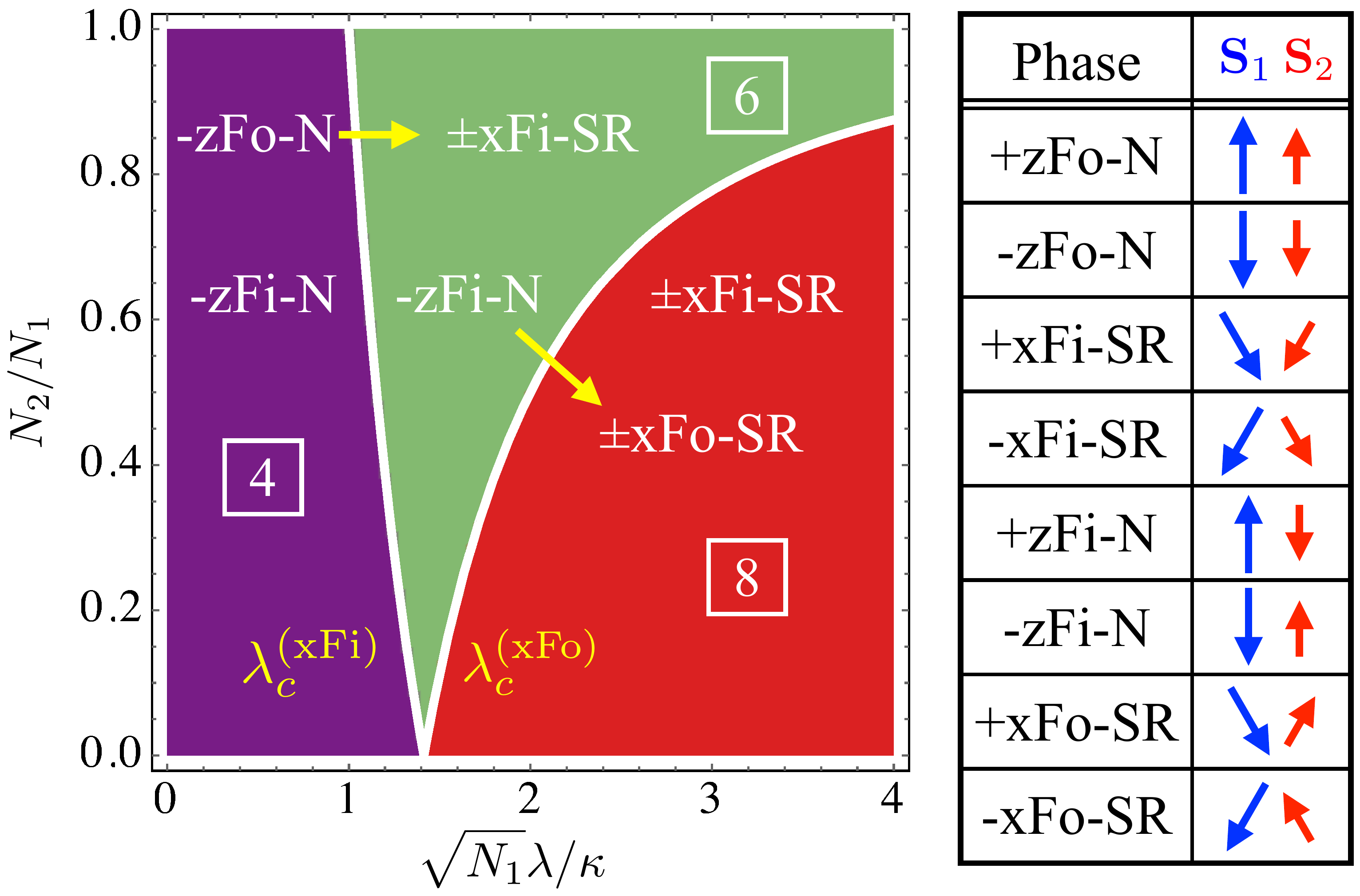}
\caption{Steady-state phase diagram of the system and schematic representations of all phases. The two white curves in the phase diagram obtained by rearranging Eq.~\eqref{eq:F/AF-threshold} indicate the thresholds for the emergence of the $\pm$xFi-SR and $\pm$xFo-SR states. The numbers inside squares show the total number of fixed points in each regime. However, only the stable solutions are stated explicitly. All the phases are illustrated schematically for finite $\lambda$ in the table in the right. The other parameters are set to $\omega_c=\omega_a=\kappa$.} 
\label{fig:ss-pd}
\end{figure}

\begin{figure}[t!]
\centering
\includegraphics [width=0.485\textwidth]{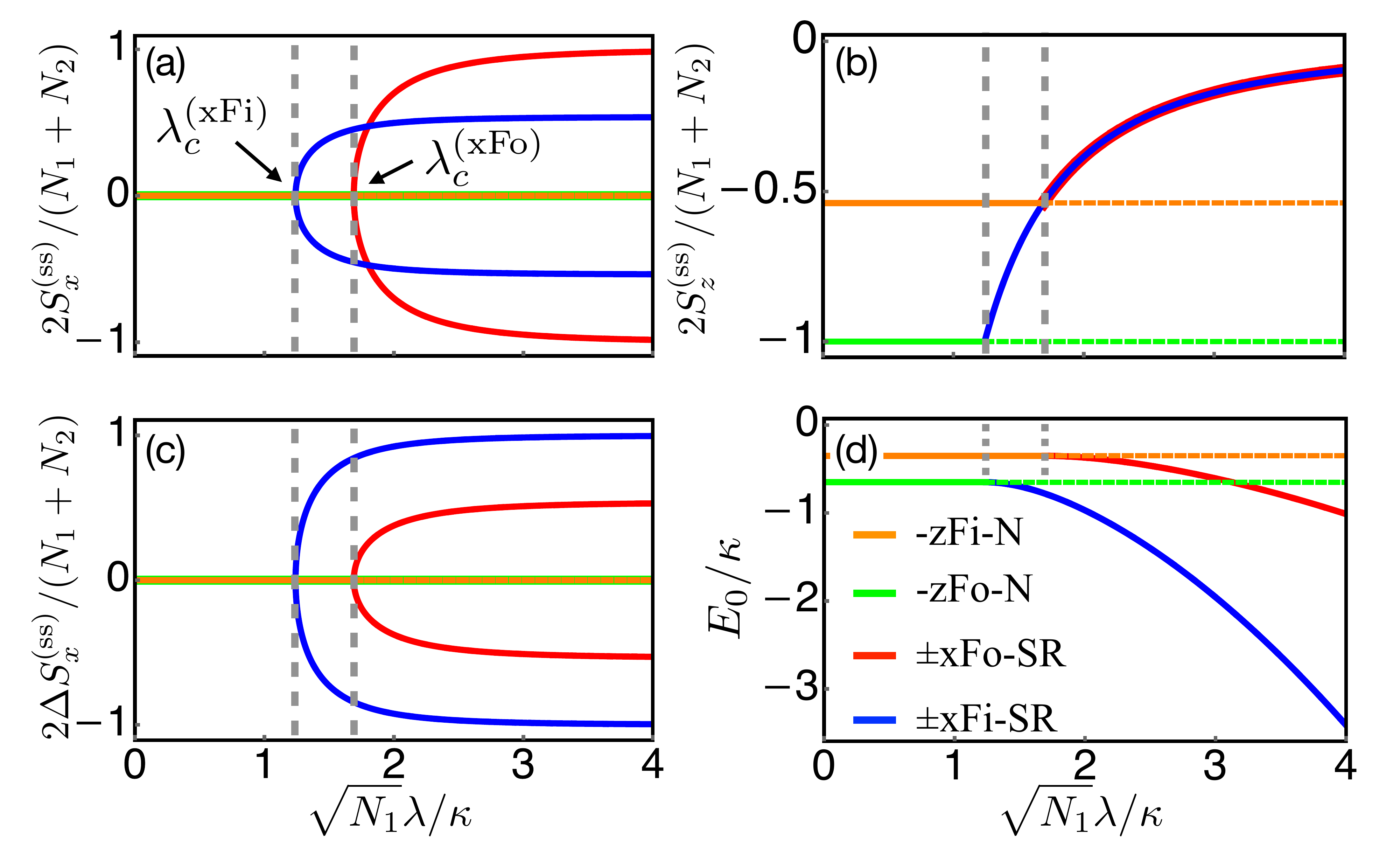}
\caption{Steady-state behavior of the system across the xFi-SR and xFo-SR thresholds: the  $\{x,z\}$-components $S_{x,z}^{(\rm ss)}$ of the total spin (a,b), the $x$-component $\Delta S_x^{(\rm ss)}$ of the staggered spin (c), and the energy $E_0$ of the system (d) as a function of the atom-field coupling $\sqrt{N_1}\lambda$ at a fixed $N_2/N_1=0.3$ for $-$zFi-N (orange), $-$zFo-N (green),  $\pm$xFo-SR (red), and $\pm$xFi-SR (blue). The $x$ component of the total and staggered spin both acquire non-zero values in the $\pm$xFi-SR and $\pm$xFo-SR states, thus implying that the system can undergo two independent  supercritical pitchfork bifurcations. The dashed green and orange lines indicate that the corresponding fixed points are unstable for the given coupling $\lambda$. The other parameters are the same as Fig.~\ref{fig:ss-pd}.
} 
\label{fig:Sxz-E0}
\end{figure}

Motivated by the recent progress in quantum-gas--cavity QED~\cite{Mivehvar2021}, in this Letter we introduce a variant of the Dicke model, which we coin the name ``\emph{non-standard} Dicke model'' for it, where two independent ensembles of spin-1/2 atoms are coupled to a single mode of a cavity with different coupling strengths. Subsequently, we focus on a special case where the coupling strengths are opposite (which is unitarily equivalent to equal-coupling strengths) and then show that the system exhibits a wealth of intriguing phenomena owing to the conservation of the total spin in each ensemble individually. In particular, the semi-classical approach reveals the existence of multistable steady-state phases as shown in Figs.~\ref{fig:ss-pd} and~\ref{fig:Sxz-E0}; especially, a bistable superradiant regime. This bistable region contains $\pm$xFo-SR states where the total spins in the both ensembles orient in the same $x$ direction (i.e., both in either positive or negative $x$ direction) in the strong coupling limit. It also includes other superradiant phases where the total spins in the two ensembles point in opposite $x$ directions (i.e., one in the positive and another in the negative $x$ directions) in the strong coupling limi, thus forming \emph{ferrimagnetic} (Fi) order ``$\pm$xFi-SR''. This order evolves into \emph{antiferromagnetic} (aF) order, ``$\pm$xaF-SR'', when two ensembles have an equal number of spins. 

\begin{figure}[t!]
\centering
\includegraphics [width=0.48\textwidth]{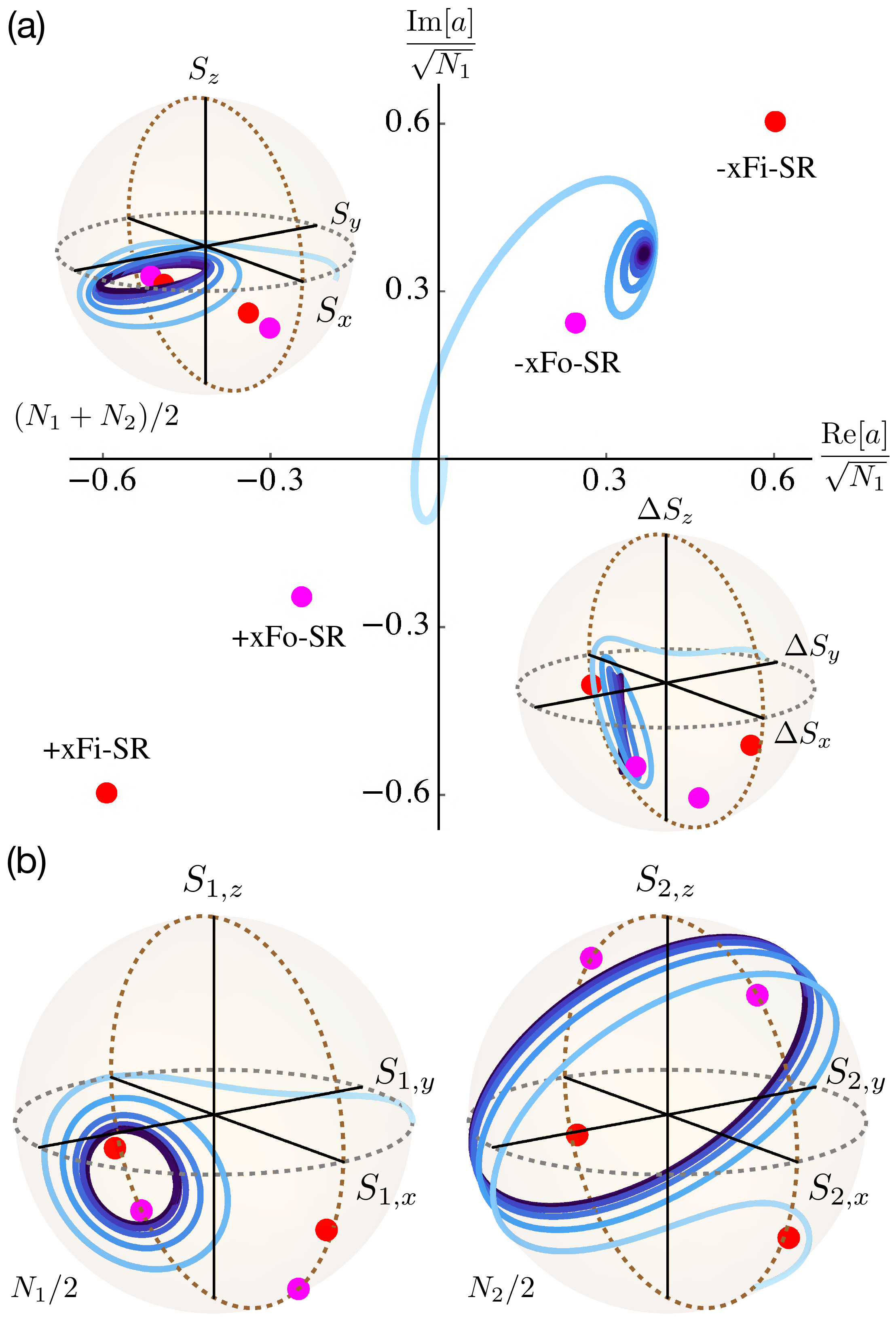}
\caption{The nonequilibrium dynamics of the system. 
The system is prepared in the initial state $\mathbf{S}_1=(1/\sqrt{2},1/\sqrt{2},0)N_1/2$, 
$\mathbf{S}_2=(1/\sqrt{2},0,-1/\sqrt{2})N_2/2$, and a small $a=\delta a$.
The ensuing dynamics of the system are displayed in the phase space of the the cavity-field
amplitude $a$ (a) and the $\mathbf{S}_1$ and $\mathbf{S}_2$ spin Bloch spheres (b).
The color gradient from light blue to darker blue in trajectories indicates schematically the arrow of the time evolution.   
The trajectories in the spin Bloch spheres exhibit limit-cycle behaviors, 
enclosing the two fixed points $-$xFo-SR and $-$xFi-SR. On the other hand, the corresponding
trajectory in the field-amplitude space is attracted to an emergent focus lying between the 
$-$xFo-SR and $-$xFi-SR fixed points. The insets in panel (a) display the dynamics in the
the Bloch spheres of the total $\mathbf{S}$ and staggered $\Delta\mathbf{S}$ spins. 
The radii of the Bloch spheres are indicated in each case, where 
$\mathbf{S}$ and $\Delta\mathbf{S}$ have the same radius $(N_1+N_2)/2$.
Here $\sqrt{N_1}\lambda/\kappa=2$ and $N_2/N_1=0.3$. 
 The other parameters are the same as Fig.~\ref{fig:ss-pd}.
} 
\label{fig:SC-dynamics-LC}
\end{figure}

\begin{figure}[t!]
\centering
\includegraphics [width=0.35\textwidth]{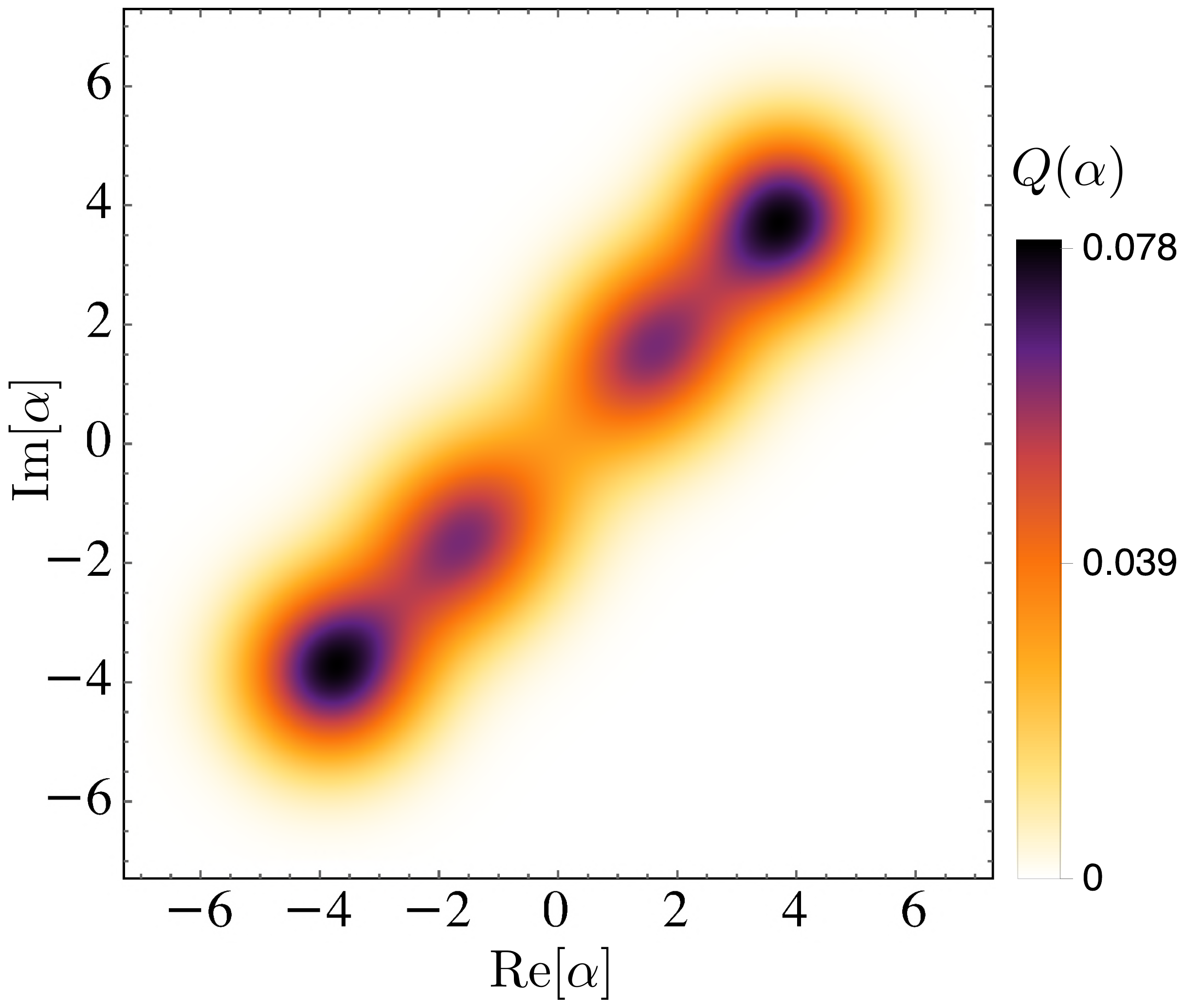}
\caption{The Husimi $Q$ representation of the cavity field
after long-time quantum dynamics starting from the initial state
$\ket{\downarrow}_1\otimes(\ket{\downarrow}_2+\ket{\uparrow}_2)/\sqrt{2}$
with $\langle\hat{\tilde{\mathbf S}}^2\rangle=36$.
It consists of four lobes, an indication of the superposition
of four $\pm$xFo-SR and $\pm$xFi-SR states.
The parameters are $N_1=10$, $N_2=4$, and $\sqrt{N_1}\lambda\simeq2.5\kappa$, with
the rest being the same as Fig.~\ref{fig:ss-pd}.
} 
\label{fig:Q-function}
\end{figure}

The linear-stability analysis as well as the nonequilibrium dynamics of the system assert the stability of the superradiant states. Furthermore, we find initial states in the multistable regimes where the ensuing nonequilibrium dynamics from them do not lead to any steady state of the system, rather give rise to non-stationary oscillating trajectories due to competing fixed points in these regimes; see Fig.~\ref{fig:SC-dynamics-LC}. Finally, the full quantum-mechanical calculations also confirm the coexistence of $\pm$xFo-SR and $\pm$xFi-SR as shown in Fig.~\ref{fig:Q-function}.

%=========================================================================
\emph{Model and Hamiltonian.}---Consider two \emph{independent} ensembles of $N_{1,2}$ spins-1/2 coupled to a single cavity mode with strengths $\lambda_{1,2}$, respectively. The system is described by a non-standard Dicke Hamiltonian,
\begin{align} \label{eq:ns-Dicke-H}
\hat{H}_{\rm nsD}=\omega_c\hat{a}^\dag\hat{a}+\omega_a\hat{S}_z
+(\hat{a}^\dag+\hat{a})(\lambda_1\hat{S}_{1,x}+\lambda_2\hat{S}_{2,x}),
\end{align}
where $\hat{S}_\alpha\equiv\hat{S}_{1,\alpha}+\hat{S}_{2,\alpha}=\sum_{j=1}^{N_1} \hat{s}_{j,\alpha}^{(1)}+\sum_{j=1}^{N_2} \hat{s}_{j,\alpha}^{(2)}$ is the total collective spin of the two ensembles. Unlike the Dicke model~\eqref{eq:Dicke-H}, the total collective spin here is not a constant of motion as $\hat{\mathbf S}^2$ does not commute with $\hat{H}_{\rm nsD}$. This means that the non-standard Dicke Hamiltonian $\hat{H}_{\rm nsD}$ mixes manifolds with different total spins~\cite{Hotter2023}. That said, the total spin $\hat{\mathbf S}^2_l$ for each ensemble is conserved. In this work, we restrict ourselves to a spacial case of $\lambda\equiv\lambda_1=-\lambda_2$. Although in this case the non-standard Dicke Hamiltonian~\eqref{eq:ns-Dicke-H} becomes unitarily equivalent to the Dicke Hamiltonian [see supplemental material (SM)~\cite{Mivehvar2023SM}], the system exhibits intriguing features due to the existence of \emph{two} individually conserved collective spins $\hat{\mathbf S}_{1}^2$ and $\hat{\mathbf S}_{2}^2$. Furthermore, in this special case another conserved quantity $\hat{\tilde{\mathbf S}}^2=(\hat{S}_{1,x}-\hat{S}_{2,x})^2+(\hat{S}_{1,y}-\hat{S}_{2,y})^2+(\hat{S}_{1,z}+\hat{S}_{2,z})^2$ emerges in the system which plays an essential role as we will see later. Due to the symmetry, in the following we restrict ourselves solely to $N_2/N_1\leq1$.

%=========================================================================
\emph{Mean-field approach: steady states, stability, and dynamics.}---In order to obtain insight into the system, we start with the mean-field approach that omits quantum fluctuations and replaces quantum operators with classical variables, namely, $\hat{a}\to a=\langle\hat{a}\rangle$ and $\hat{\mathbf S}_{l}\to \mathbf S_{l}=\langle\hat{\mathbf S}_{l}\rangle$ with $l=1,2$, justified for large ensembles of spins~\cite{Keeling2010Collective, Bhaseen2012Dynamics, Carollo2021}. The system is then described by a set of seven coupled differential equations obtained from Heisenberg equations of motion,
\begin{align} \label{eq:sc-eq-motion}
\dot{a}&=-(i\omega_c+\kappa)a-i\lambda(S_{1,x}-S_{2,x}),\nonumber\\
\dot{S}_{l,x}& = -\omega_a S_{l,y}, 
\nonumber\\
\dot{S}_{l,y}& = \omega_a S_{l,x} + (-1)^l \lambda (a^*+a)S_{l,z}, 
\nonumber\\
\dot{S}_{l,z}& = (-1)^{l+1} \lambda (a^*+a)S_{l,y},
\end{align}
and endowed by \emph{two} spin-conservation constraints $S_l=|{\mathbf S}_l|=\sqrt{S_{l,x}^2+S_{l,y}^2+S_{l,z}^2}=N_l/2$. Here, $\kappa$ is the cavity-field decay rate.

The fixed-point solutions are obtained by setting $\dot{a}=0$ and $\dot{S}_{l,\alpha}=0$ in Eq.~\eqref{eq:sc-eq-motion}. This immediately implies that in a steady state, $S_{l,y}^{\rm (ss)}=0$ for any $\omega_a\neq0$. After some algebraic manipulation~\cite{Mivehvar2023SM}, one obtains two distinct classes of nontrivial solutions: 
\begin{subequations} \label{eq:ss-Slx}
\begin{align} \label{eq:ss-Slx-xFo}
S_{1,x}^{\rm (ss)}=\frac{N_1}{N_2}S_{2,x}^{\rm (ss)}=
\pm \frac{N_1}{2} 
\sqrt{1-\bigg(\frac{\lambda_c^{(\rm xFo)}}{\lambda}\bigg)^4},
\end{align}
with corresponding $S_{l,z}^{\rm (ss)}=(-1)^l\sqrt{(N_l/2)^2-[S_{l,x}^{\rm (ss)}]^2}$ determined through the normalization constraints, and
\begin{align} \label{eq:ss-Slx-xFi}
S_{1,x}^{\rm (ss)}=-\frac{N_1}{N_2}S_{2,x}^{\rm (ss)}=
\pm\frac{N_1}{2} 
\sqrt{1-\bigg(\frac{\lambda_c^{(\rm xFi)}}{\lambda}\bigg)^4},
\end{align}
\end{subequations} 
with corresponding $S_{l,z}^{\rm (ss)}=-\sqrt{(N_l/2)^2-[S_{l,x}^{\rm (ss)}]^2}$. Here, we have introduced the critical couplings as
\begin{align} \label{eq:F/AF-threshold}
\lambda_c^{(\rm xFo/xFi)}=\sqrt{\frac{\omega_a(\omega_c^2+\kappa^2)}{(N_1\mp N_2)\omega_c}}. 
\end{align}
The choice of the sign for $S_{l,x}^{\rm (ss)}$, i.e., the ``$\pm$'' sings outside of the square roots in the right-hand sides of Eq.~\eqref{eq:ss-Slx}, reflects two possible $\mathbb{Z}_2$ solutions originating from the invariance of the Hamiltonian~\eqref{eq:ns-Dicke-H} under the parity transformation $\hat{a}\to-\hat{a}$ and $\hat{S}_{l,x}\to -\hat{S}_{l,x}$.

%--------Figure------------ 

The two nontrivial classes of the solutions, Eqs.~\eqref{eq:ss-Slx-xFo} and~\eqref{eq:ss-Slx-xFi}, have fundamentally different properties. The first class corresponds to superradiant states with $x$-ferromagnetic ordering where both $S_{1,x}^{\rm (ss)}$ and $S_{2,x}^{\rm (ss)}$ point in the same direction, while the second class corresponds to superradiant states with $x$-ferrimagnetic ordering where $S_{1,x}^{\rm (ss)}$ and $S_{2,x}^{\rm (ss)}$ point in the opposite directions. We refer to these superradiant states, respectively, as ``$\pm$xFo-SR'' and ``$\pm$xFi-SR'', corresponding to their asymptotic spin-ordering behaviors at $\lambda\to\infty$; see the table in the right side of Fig.~\ref{fig:ss-pd} for the schematic representation of these states at finite $\lambda$. The $\pm$xFi-SR states cross over into $x$-antiferromagnetic superradiant states, ``$\pm$xaF-SR'', for $N_1=N_2$. 

In addition to the four superradiant fixed points discussed above, there are four trivial fixed points corresponding to $a_{\rm ss}=S_{l,x}^{\rm (ss)}=0$ and either $S_{z}^{\rm (ss)}=\pm (N_1+N_2)/2$ or $S_{z}^{\rm (ss)}=\pm (N_1- N_2)/2$. These states are designated, respectively, by ``$\pm$zFo-N'' and ``$\pm$zFi-N''; see the table in the right side of Fig.~\ref{fig:ss-pd}. Once again, the $\pm$zFi-N states cross over into $z$-antiferromagnetic normal states, ``$\pm$zaF-N'', for $N_1=N_2$.

The superradiant thresholds are obtained by setting $S_{l,x}^{\rm (ss)}=0$ in Eq.~\eqref{eq:ss-Slx}, yielding $\lambda=\lambda_c^{(\rm xFo/xFi)}$. The fixed points $\pm$xFo-SR ($\pm$xFi-SR) only emerge beyond the threshold $\lambda_c^{(\rm xFo)}$ [$\lambda_c^{(\rm xFi)}$]. Note that the xFo-SR threshold diverges for $N_2\to N_1$ as $\alpha_{\rm ss}\propto[{S}_{1,x}^{\rm (ss)}-{S}_{2,x}^{\rm (ss)}]\to0$. 

We now turn our attention to the linear stability of the fixed points of the system obtained above. To this end, we write $a=a_{\rm ss}+\delta a$ and $S_{l,\alpha}=S_{l,\alpha}^{\rm (ss)}+\delta S_{l,\alpha}$ in the mean-field equations of motion~\eqref{eq:sc-eq-motion} and subsequently linearize them to obtain
$\partial_t\delta\mathbf{X}=\mathbf{J} \delta\mathbf{X}$,
where $\delta\mathbf{X}$ is a vector of fluctuations $\delta a$ and $\delta S_{l,\alpha}$, and $\mathbf{J}$ the Jacobian matrix given explicitly in SM~\cite{Mivehvar2023SM}. A fixed point is stable provided all eigenvalues of the Jacobian matrix for that given fixed point have negative real parts~\cite{Roussel2019}. We find that all $\pm$xFo-SR and $\pm$xFi-SR states are stable in their entire corresponding parameter regimes, i.e., beyond the $\lambda_c^{(\rm xFo/xFi)}$ thresholds, respectively. On the other hand, the trivial fixed points $-$zFo-N and $-$zFi-N are solely stable below the $\lambda_c^{(\rm xFi)}$ and $\lambda_c^{(\rm xFo)}$ thresholds, respectively, and lose their stability beyond these thresholds. The other two trivial fixed points +zFo-N and +zFi-N are always unstable. This implies the system undergoes two independent supercritical pitchfork bifurcations from $-$zFo-N into $\pm$xFi-SR at the threshold $\lambda_c^{(\rm xFi)}$, and from $-$zFi-N into $\pm$xFo-SR at the threshold $\lambda_c^{(\rm xFo)}$. The conservation of ${\tilde{\mathbf S}}^2=\langle\hat{\tilde{\mathbf S}}^2\rangle$ provides an account for the independence of these two bifurcations: ${\tilde{\mathbf S}}^2=(N_1+N_2)^2/4$ in both $-$zFo-N and $\pm$xFi-SR states, while it is equal to $(N_1-N_2)^2/4$ in $-$zFi-N and $\pm$xFo-SR states. Therefore, these two phase transitions lie in different symmetry sectors and are independent from each other.

The steady-state phase diagram of the system in the parameter plane of $\{N_2/N_1,\sqrt{N_1}\lambda/\kappa\}$ is shown in Fig.~\ref{fig:ss-pd}. The two white curves are the analytical boundaries for the xFo-SR and xFi-SR transitions, obtained from Eq.~\eqref{eq:F/AF-threshold}. The total number of fixed points in each parameter regime is indicated in the phase diagram. However, only the stable fixed points in each parameter regime are indicated explicitly. The phase diagram features regions of multistability~\cite{Gabor2023Ground}, in particular, a region with multiple coexistent superradiant phases, reminiscent of optical bistability due to optomechanical effects in longitudinally driven atom-cavity systems~\cite{Gupta2007, Brennecke2008, Larson2008, Zhou2009, Ali2022}.

%--------Figure------------ 

The steady-state behavior of the $\{x,z\}$-components of the total spin $S_{x,z}^{(\rm ss)}$, the $x$-component of the staggered spin $\Delta{S}_x^{(\rm ss)}\equiv S_{1,x}^{(\rm ss)}-S_{2,x}^{(\rm ss)}$, and the energy of the system $E_0=\langle \hat{H}_{\rm nsD}\rangle$ as a function of the atom-field coupling $\lambda$ at a fixed $N_2/N_1=0.3$ are depicted in Fig.~\ref{fig:Sxz-E0} for the fixed points $-$zFo-N, $-$zFi-N, $\pm$xFo-SR, and $\pm$xFi-SR. As expected,  $\Delta{S}_x^{(\rm ss)}$ acquires nonzero values across the $\lambda_c^{(\rm xFi/xFo)}$ thresholds in the $\pm$xFi-SR (blue curves) and $\pm$xFo-SR (red curves) states, respectively; see Fig.~\ref{fig:Sxz-E0}(c). The two branches in each case correspond to the two possible $\mathbb{Z}_2$ solutions. Accordingly, the field amplitude $\alpha_{\rm ss}\propto\Delta{S}_x^{(\rm ss)}$ also grows from zero and the system enters superradiant phases, implying that $\Delta{S}_x^{(\rm ss)}$ can be identified as the order parameter of the system. Note that $\Delta{S}_x^{(\rm ss)}$ is zero in both $-$zFo-N and $-$zFi-N states [green and orange lines, respectively; obstructed somewhat by each other in Fig.~\ref{fig:Sxz-E0}(c)]. The $x$ component of the total spin, $S_{x}^{(\rm ss)}$, exhibits a similar behavior, though it grows much faster in the $\pm$xFo-SR states compared to the $\pm$xFi-SR states; see Fig.~\ref{fig:Sxz-E0}(a). From Fig.~\ref{fig:Sxz-E0}(b) and~(d), one sees that the $z$-component $S_z^{(\rm ss)}$ of the total spin and the energy $E_0$ change continuously from the $-$zFo-N to $\pm$xFi-SR states and from the $-$zFi-N into $\pm$xFo-SR states, signaling second-order phase transitions. This is consistent with Fig.~\ref{fig:Sxz-E0}(c) and the linear-stability analysis predicting supercritical pitchfork bifurcations. The behavior of $S_{x,z}^{(\rm ss)}$, $\Delta{S}_x^{(\rm ss)}$, and $E_0$ in the full parameter space of $\{N_2/N_1,\sqrt{N_1}\lambda/\kappa\}$ are given in SM~\cite{Mivehvar2023SM}.

We now examine the semiclassical dynamics of the system by numerically integrating the mean-field equations of motion~\eqref{eq:sc-eq-motion}. As expected, the stable steady states are the attractors of the long-time dynamics, while a small perturbation in the unstable steady states leads to phase-space trajectories being repelled from these points into one of the stable fixed points. We study these by adding a small perturbation $\delta a$ to the steady-state field amplitude $a_{\rm ss}$ in each fixed point and looking at the ensuing dynamics which is strictly constrained by the conservation of ${\tilde{\mathbf S}}^2$. In particular, this reveals that both $\pm$zFo-N ($\pm$zFi-N) are unstable towards one of the parity-symmetric pair $\pm$xFi-SR ($\pm$xFo-SR) above the threshold $\lambda_c^{(\rm xFi)}$ [$\lambda_c^{(\rm xFo)}$]; see SM for these types of phase-space dynamics. While below the threshold, +zFo-N (+zFi-N) evolves to the corresponding low-energy state $-$zFo-N ($-$zFi-N).

Note that due to the multistability, the long-time dynamics depend crucially on the initial state and can exhibit intriguing features. In particular, in some parameter regimes we find initial states where the following phase-space dynamics from them do not lead to any of the stable fixed points, rather exhibit oscillatory behaviors owing to competing fixed points~\cite{Thompson1984}. This is especially interesting in the parameter regime where all $\pm$xFo-SR and $\pm$xFi-SR are the stable fixed points of the system. Starting from the initial state $\mathbf{S}_1=(1/\sqrt{2},1/\sqrt{2},0)N_1/2$ and $\mathbf{S}_2=(1/\sqrt{2},0,-1/\sqrt{2})N_2/2$ with a small field $a=\delta a$, the ensuing trajectories in the Bloch spheres of $\mathbf{S}_1$ and $\mathbf{S}_2$ encircle the two fixed points $-$xFo-SR and $-$xFi-SR and exhibit limit cycles as shown in Fig.~\ref{fig:SC-dynamics-LC}(b)~\cite{Kosior2023}. That said, the corresponding trajectory in the phase space of the cavity-field amplitude $a$ as shown in Fig.~\ref{fig:SC-dynamics-LC}(a) exhibits an inward spiral behavior toward an emergent focus lying between the two fixed points $-$xFo-SR and $-$xFi-SR. These nonequilibrium dynamics can be understood by noting that the initial state with ${\tilde{\mathbf S}}^2=(N_1^2+N_2^2)/8$ lies in a different symmetry sector than the xFo-SR and xFi-SR states. The dynamics of the system in the Bloch spheres of the total $\mathbf{S}$ and staggered $\Delta\mathbf{S}$ spins with radius $(N_1+N_2)/2$ are displayed in the insets of  Fig.~\ref{fig:SC-dynamics-LC}(a). Note that although the dynamics in the Bloch spheres $\mathbf{S}_1$ and $\mathbf{S}_2$ are strictly restricted to the Bloch surfaces, in the $\mathbf{S}$ and $\Delta\mathbf{S}$ phase spaces the dynamics move inside the Bloch spheres, confirming that the total spin $\mathbf{S}$ is not conserved.

%--------Figure------------ 

%=========================================================================
\emph{Quantum description.}---Finally, we address briefly the full quantum description of the system via the master equation for the density-matrix operator, $\hat{\dot\rho}=i[\hat{\rho},\hat{H}_{\rm nsD}]+\hat{\mathcal{L}}[\hat{\rho}]$. In the Born-Markov approximation, the Liouvillian can be expressed in the Lindblad form as $\hat{\mathcal{L}}[\hat{\rho}]=\kappa(2\hat{a}\hat{\rho}\hat{a}^\dag-\hat{a}^\dag\hat{a}\hat{\rho}-\hat{\rho}\hat{a}^\dag\hat{a})$, where we have ignored the decay of the atomic excited states.

We perform small-scale, full-quantum calculations with $N_1=10$ and $N_2=4$ and find that indeed depending on the initial state, $\pm$xFo-SR or $\pm$xFi-SR can be the final state of the long-time quantum dynamics of the system in appropriate parameter regimes; see SM~\cite{Mivehvar2023SM}. Remarkably, we also find initial states and parameter regimes where the ensuing quantum dynamics from them lead to a state which is a superposition of all $\pm$xFo-SR and $\pm$xFi-SR. This can be best seen in the $Q$-representation of the cavity field~\cite{Sandner2015,Halati2020}. The $Q$-function is presented in Fig.~\ref{fig:Q-function} and comprised of four, partially disjoint lobes, an indication of the coexistence of multiple superradiant states, i.e., $\pm$xFo-SR and $\pm$xFi-SR [cf.~Fig.~\ref{fig:SC-dynamics-LC}(a)]. 

%--------Figure------------ 

%=========================================================================
\emph{Conclusions.}---We have introduced a variant of the Dicke model where two independent ensembles of spins couple to a single cavity mode with different coupling strengths $\lambda_{1,2}$. When $\lambda_1=-\lambda_2$, the Hamiltonian maps unitarily to the Dicke model. That said, the system still exhibits intriguing steady-state and non-stationary phenomena owing to the conservation of the total spin in each ensemble, effectively allowing one to explore physics beyond the totally symmetric Dicke subspace. In the general case when $\lambda_1\neq\pm\lambda_2$, the dynamics of the system will be richer and more complex since $\tilde{\mathbf S}^2$ is no longer a constant of motion and different symmetry sectors mix with one another. As a further consequence, the system might exhibit a different critical behavior and may possess multicritical points~\cite{Soriente2018, Xu2019, Zhu2020}. These aspects will be considered in a future work and presented elsewhere. Our model can readily be implemented in state-of-the-art experiments~\cite{Gupta2007, Muniz2020, Bohr2023Collectively} as discussed in SM~\cite{Mivehvar2023SM} and opens a new avenue for studying various nonequilibrium magnetic ordering~\cite{Davis2019} and dynamical phenomena~\cite{Dogra2019,Kongkhambut2022} in cavity-QED experimental setups.

%=========================================================================
\begin{acknowledgments}
I acknowledge inspiring discussions with Natalia Masalaeva, Karol Gietka, Arkadiusz Kosior, Christoph Hotter, and Helmut Ritsch. I am also grateful to Jonathan Keeling, Brendan Marsh, and Benjamin Lev for enlightening and fruitful communications.
This research was funded in whole or in part by the Austrian Science Fund (FWF) [grant DOI: 10.55776/P35891]. For open access purposes, the author has applied a CC BY public copyright license to any author accepted manuscript version arising from this submission.
F.\,M.~is also supported financially by the Tyrolean Science Promotion Fund (TWF) and the ESQ-Discovery Grant of the Austrian Academy of Sciences (\"OAW).
\end{acknowledgments}

%=========================================================================
\bibliography{NSD}

%=========================================================================
\widetext
\newpage
\setcounter{equation}{0}
\setcounter{figure}{0}
\renewcommand{\theequation}{S\arabic{equation}}
\renewcommand{\thefigure}{S\arabic{figure}}

%=========================================================================
\section{Supplemental Material}
In the supplemental material, we present the details of the unitary equivalence of the non-standard Dicke model in the special case of $\lambda_1=-\lambda_2$ to the Dicke model, the analytical derivation of the steady states, their linear-stability analysis, the dependence of the semi-classical and quantum dynamics of the system on the initial state, and a possible experimental implementation.

%=========================================================================
\subsection{The unitary equivalence of the non-standard Dicke model in the spacial case of $\lambda_1=-\lambda_2$ to the Dicke model}
\label{sec:unitary-equivalence}

In the special case when $\lambda\equiv\lambda_1=-\lambda_2$, the non-standard Dicke Hamiltonian~\eqref{eq:ns-Dicke-H} takes a simple form, 
\begin{align} \label{eq:SM-ns-Dicke-H-special}
\hat{H}_{\rm nsD}=\omega_c\hat{a}^\dag\hat{a}
+\omega_a(\hat{S}_{1,z}+\hat{S}_{2,z})
+\lambda(\hat{a}^\dag+\hat{a})(\hat{S}_{1,x}-\hat{S}_{2,x}).
\end{align}
By rotating the second collective spin around $\hat{S}_{2,z}$ by $\pi$, described by the unitary transformation $\hat{\mathcal U}=e^{-i\pi\hat{S}_{2,z}}$, one obtains $\hat{\mathbf S}_2 \to \hat{\mathcal U}^\dag\hat{\mathbf S}_2\hat{\mathcal U}=(-\hat{S}_{2,x},-\hat{S}_{2,y},\hat{S}_{2,z})$. Consequently, the non-standard Dicke model in the special case of $\lambda\equiv\lambda_1=-\lambda_2$, Eq.~\eqref{eq:SM-ns-Dicke-H-special}, transforms into
\begin{align} \label{eq:SM-ns-Dicke-H-special-rotated}
\hat{\tilde H}_{\rm nsD}= 
\hat{\mathcal U}^\dag \hat{H}_{\rm nsD} \hat{\mathcal U} = 
\omega_c\hat{a}^\dag\hat{a}
+\omega_a(\hat{S}_{1,z}+\hat{S}_{2,z})
+\lambda(\hat{a}^\dag+\hat{a})(\hat{S}_{1,x}+\hat{S}_{2,x}),
\end{align}
which is unitarily equivalent to the Dicke model. The interesting phenomena established in this work originates from the fact that, there are two independent spin ensembles and the collective spin of each ensemble is conserved independently. Even by preparing each ensemble in its totally symmetric subspace where $S_l=N_l/2$, the spin (i.e., angular momentum) addition rule implies that the total collective-spin quantum number can take any allowed value in the interval $|N_1-N_2|/2,\cdots,(N_1+N_2)/2$. Therefore, the relative orientation of the two collective spins provides an efficient way to prepare the system in spin manifolds beyond the totally symmetric subspace. 

The quantity $\hat{\tilde{\mathbf S}}^2$ defined in the main text has a more clear meaning in the rotated frame: $\hat{\mathcal U}^\dag\hat{\tilde{\mathbf S}}^2\hat{\mathcal U}=(\hat{S}_{1,x}+\hat{S}_{2,x})^2+(\hat{S}_{1,y}+\hat{S}_{2,y})^2+(\hat{S}_{1,z}+\hat{S}_{2,z})^2$. It is just the square of the total collective spin in the rotated frame. Note that the steady states are also transformed unitarily via $\hat{\mathcal U}$; see Table~\ref{tab:SM_transformed_phases} for the schematic representation of the fixed points in the rotated frame. 

\begin{table}[b!]
   \centering
      \includegraphics[width=0.8\textwidth]{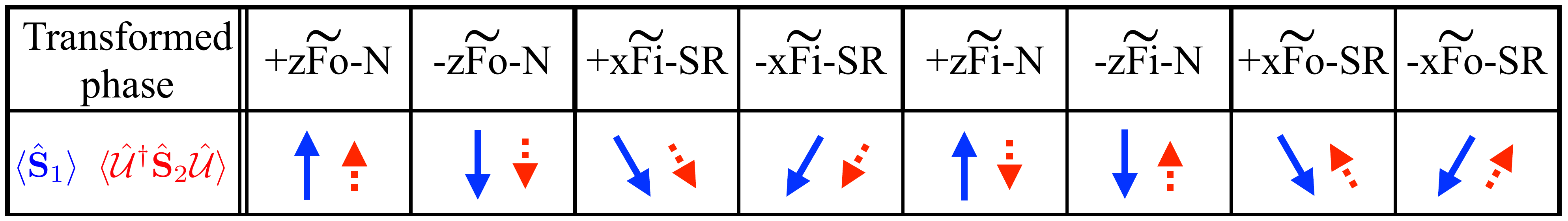}
      \caption{Schematic representation of all the mean-field fixed points of the system in the transformed frame; cf.\ Fig.~\ref{fig:ss-pd} in the main text.
      The second collective spin is rotated around $\hat{S}_{2,z}$ by $\pi$ according to 
      $\hat{\mathbf S}_2\to \hat{\mathcal U}^\dag\hat{\mathbf S}_2\hat{\mathcal U} =(-{S}_{2,x},-{S}_{2,y},{S}_{2,z})$. The rotated $\hat{\mathbf S}_2$
      is represented schematically by the dashed arrow.}
   \label{tab:SM_transformed_phases}
\end{table}

%=========================================================================
\subsection{Steady-state solutions} 
\label{sec:SM-ss}

In this section we derive the steady-state solutions in more details. By setting $\dot{a}=0$ and $\dot{S}_{l,\alpha}=0$ in Eq.~\eqref{eq:sc-eq-motion} in the main text, one obtains
\begin{subequations}
\begin{align} \label{SM-eq:a'=0}
a_{\rm ss}&=\frac{\lambda}{-\omega_c+i\kappa}\left[S_{1,x}^{\rm (ss)}-S_{2,x}^{\rm (ss)}\right],
\end{align}
\begin{align} \label{SM-eq:Slx'=0}
S_{l,x}^{\rm (ss)}&=(-1)^{l+1}\frac{\lambda}{\omega_a}(a_{\rm ss}+a_{\rm ss}^*)S_{l,z}^{\rm (ss)} \quad \text{with} \quad l=1,2.
\end{align}
\end{subequations}
Dividing the two equations in Eq.~\eqref{SM-eq:Slx'=0} leads to a relation between $S_{1,x/z}^{\rm (ss)}$ and $S_{2,x/z}^{\rm (ss)}$,
\begin{align} \label{eq:SM-ss-Slx-Slz-relation}
\frac{S_{1,x}^{\rm (ss)}}{S_{2,x}^{\rm (ss)}}=-\frac{S_{1,z}^{\rm (ss)}}{S_{2,z}^{\rm (ss)}}.
\end{align}
Here, $S_{1,z}^{\rm (ss)}$ and $S_{2,z}^{\rm (ss)}$ can have the same or opposite signs, which lead to two distinct classes of nontrivial solutions. In particular, when $S_{1,z}^{\rm (ss)}$ and $S_{2,z}^{\rm (ss)}$ have the same sign (opposite signs), $S_{1,x}^{\rm (ss)}$ and $S_{2,x}^{\rm (ss)}$ must have opposite signs (the same sign), in order for Eq.~\eqref{eq:SM-ss-Slx-Slz-relation} to be satisfied. These correspond, respectively, to the $\pm$xFi-SR and $\pm$xFo-SR ordering. 

By noting that in the steady states $|S_{l,z}^{\rm (ss)}|=\sqrt{(N_l/2)^2-[S_{l,x}^{\rm (ss)}]^2}$, Eq.~\eqref{eq:SM-ss-Slx-Slz-relation} can be expressed in terms of only  $S_{l,x}^{\rm (ss)}$ as,
\begin{align} \label{eq:SM-ss-Slx-relation}
\frac{S_{1,x}^{\rm (ss)}}{S_{2,x}^{\rm (ss)}}
=\pm\frac{\sqrt{\left(\frac{N_1}{2}\right)^2-[S_{1,x}^{\rm (ss)}]^2}}{\sqrt{\left(\frac{N_2}{2}\right)^2-[S_{2,x}^{\rm (ss)}]^2}}.
\end{align}
Equation~\eqref{eq:SM-ss-Slx-relation} can be simplified to yield the relation $S_{1,x}^{\rm (ss)}=\pm(N_1/N_2)S_{2,x}^{\rm (ss)}$, where the upper plus (lower minus) sign corresponds to the xFo-SR (xFi-SR). Using this relation, the steady-state field amplitude $a_{\rm ss}$ [Eq.~\eqref{SM-eq:a'=0}] can be expressed in terms of one of the ensemble spins, say $S_{1,x}^{\rm (ss)}$. Substituting this steady-sate field amplitude back in the corresponding equation for the spin $S_{1,x}^{\rm (ss)}$ [Eq.~\eqref{SM-eq:Slx'=0}], one obtains 
\begin{align} \label{SM-eq:Slx'=0-no-a}
1&=-\frac{2\omega_c\lambda^2}{\omega_a(\omega_c^2+\kappa^2)}
\left(1\mp\frac{N_2}{N_1}\right)
\sqrt{\left(\frac{N_1}{2}\right)^2-\left[S_{1,x}^{\rm (ss)}\right]^2}.
\end{align}
This equation can readily be solved for $S_{1,x}^{\rm (ss)}$ and subsequently $S_{2,x}^{\rm (ss)}$, 
\begin{align} \label{eq:SM-ss-Slx}
S_{1,x}^{\rm (ss)}=
\pm\sqrt{\left(\frac{N_1}{2}\right)^2-\left[\frac{\omega_a(\omega_c^2+\kappa^2)}{2\left(1\mp\frac{N_2}{N_1}\right)\omega_c\lambda^2}\right]^2},\nonumber\\
S_{2,x}^{\rm (ss)}=
\pm\sqrt{\left(\frac{N_2}{2}\right)^2-\left[\frac{\omega_a(\omega_c^2+\kappa^2)}{2\left(\frac{N_1}{N_2}\mp1\right)\omega_c\lambda^2}\right]^2},
\end{align}
yielding, after defining the critical couplings $\lambda_c^{(\rm xFo/xFi)}=\sqrt{\omega_a(\omega_c^2+\kappa^2)/(N_1\mp N_2)\omega_c}$, Eq.~\eqref{eq:ss-Slx} in the main text.

After obtaining $S_{l,x}^{\rm (ss)}$, the steady-state $S_{l,z}^{\rm (ss)}$ can be obtained from the normalization constraints, $S_{l,z}^{\rm (ss)}=\pm\sqrt{(N_l/2)^2-[S_{l,x}^{\rm (ss)}]^2}$. As discussed above, $S_{1,z}^{\rm (ss)}$ and $S_{2,z}^{\rm (ss)}$ have the same sign (opposite signs) in the $\pm$xFi-SR ($\pm$xFo-SR). In order to determine the exact signs, however, we numerically solve the steady-state equations. We find that in the $\pm$xFi-SR states the minus sign is always picked up for both $S_{l,z}^{\rm (ss)}$, that is, $S_{l,z}^{\rm (ss)}=-\sqrt{(N_l/2)^2-[S_{l,x}^{\rm (ss)}]^2}$. On the other hand, for the $\pm$xFo-SR states the minus sign is picked up for $S_{1,z}^{\rm (ss)}$ and the plus sign for $S_{2,z}^{\rm (ss)}$, i.e., $S_{l,z}^{\rm (ss)}=(-1)^l\sqrt{(N_l/2)^2-[S_{l,x}^{\rm (ss)}]^2}$. Indeed, the linear-stability analyses confirm that with these choices, all $\pm$xFi-SR and $\pm$xFo-SR are stable in their entire corresponding parameter regimes, while the other choices lead to instabilities in some parameter regimes. Furthermore, Figs.~\ref{fig:Sxz-E0}(b) and (d) also show that with these choices, the $z$-component $S_{z}^{\rm (ss)}$ of the total spin and the energy $E_0$ of the system change continuously between the $-$zFo-N and $\pm$xFi-SR states, and the $-$zFi-N and $\pm$xFo-SR states across the phase transitions.

%--------Figure------------ 
\begin{figure}[t!]
\centering
\includegraphics [width=0.99\textwidth]{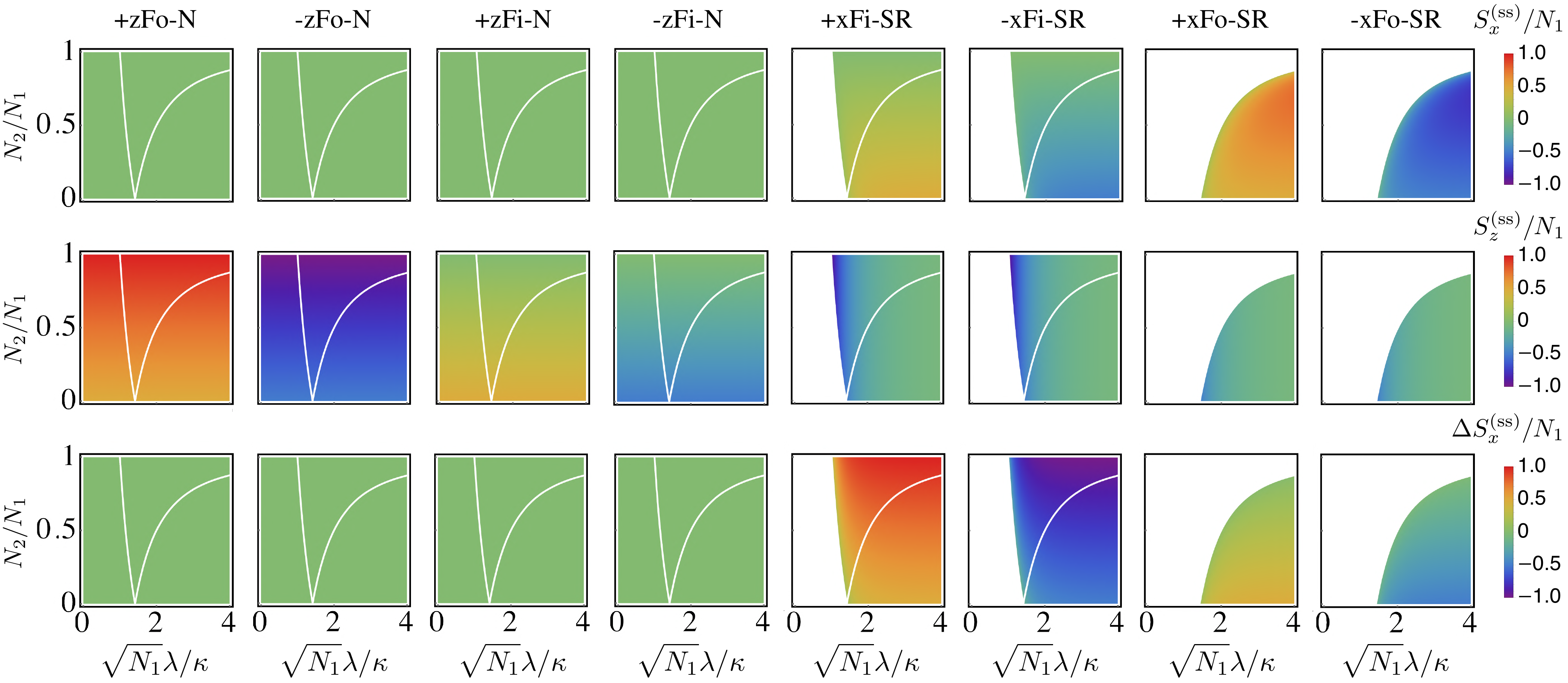}
\caption{The steady-state behavior of the $x$ and $z$ components of the total spin, $S^{(\rm ss)}_{x}$ and $S^{(\rm ss)}_{z}$, and the $x$ component of the staggered spin, $\Delta S^{(\rm ss)}_x$, in the parameter plan of  $N_2/N_1$ vs.\ $\sqrt{N_1}\lambda/\kappa$, similar to the phase diagram of Fig.~\ref{fig:ss-pd} in the main text. Each column shows the behavior 
of a fixed point which is indicated in the top. The white areas in the figures for the fixed
points $\pm$xFi-SR and $\pm$xFo-SR indicate parameter regimes where the corresponding fixed
points do not exist. The parameters are the same as Fig.~\ref{fig:ss-pd} in the main text. 
} 
\label{fig_SM:Sx_SSx_Sz}
\end{figure}

In Fig.~\ref{fig_SM:Sx_SSx_Sz}, we show the steady-state behavior of the $x$ and $z$ components of the total spin, $S_{x}^{(\rm{ss})}$ and $S_{z}^{(\rm{ss})}$, and the $x$ component of the staggered spin, $\Delta S_x^{(\rm{ss})}$, in the parameter plan of  $N_2/N_1$ vs.\ $\sqrt{N_1}\lambda/\kappa$ in accordance with the phase diagram of Fig.~\ref{fig:ss-pd} in the main text. Note that +xFi-SR and $-$xFi-SR, and +xFo-SR and $-$xFo-SR are the parity-symmetric pairs, related to one another by the parity transformation $a\to-a$ and $S_{l,x}\to-S_{l,x}$ (and equivalently $S_{x}\to-S_{x}$ and $\Delta S_{x}\to-\Delta S_{x}$).

%=========================================================================
\subsection{Linear-stability analysis}

We begin by writing the mean-field equations of motion~\eqref{eq:sc-eq-motion} in the form,
\begin{align} \label{SM-eq:sc-eq-motion-compact}
\frac{\partial\mathbf{X}}{\partial t}=\mathbf{F}(\mathbf{X}),
\end{align}
where $\mathbf{X}=(a,a^*,\mathbf{S}_1,\mathbf{S}_2)^\top$ is a vector of the mean-field variables and $\mathbf{F}(\mathbf{X})$ an $8\times 8$ matrix obtained from the right-hand side of Eq.~\eqref{eq:sc-eq-motion}. The fixed points $\dot{\mathbf X}_{\rm ss}=0$ of the system are obtained by solving the coupled algebraic equations $\mathbf{F}(\mathbf{X}_{\rm ss})=0$, as discussed in the previous section. Taking a Taylor expansion (up to the linear term) of the right-hand side of Eq.~\eqref{SM-eq:sc-eq-motion-compact} around the fixed point $\mathbf{X}_{\rm ss}$ and noting that $\partial_t{\mathbf X}_{\rm ss}=\mathbf{F}(\mathbf{X}_{\rm ss})=0$ yields~\cite{Roussel2019},
\begin{align}
\frac{\partial}{\partial t}\delta\mathbf{X}=\mathbf{J} \delta\mathbf{X},
\end{align}
where $\delta\mathbf{X}=\mathbf{X}-\mathbf{X}_{\rm ss}=(\delta a,\delta a^*,\delta \mathbf{S}_1,\delta \mathbf{S}_2)^\top$ is a vector of fluctuations, and $\mathbf{J}$ the Jacobian matrix
\begin{align}
\mathbf{J}=\frac{\partial\mathbf{F}(\mathbf{X})}{\partial\mathbf{X}}\Bigg|_{\mathbf{X}_{\rm ss}}=
\begin{pmatrix}
-(i\omega_c+\kappa) & 0 & -i\lambda & 0 & 0 & i\lambda & 0 & 0 \\
0 & -(-i\omega_c+\kappa) & i\lambda & 0 & 0 & -i\lambda & 0 & 0 \\
0 & 0 & 0 & -\omega_a & 0 & 0 & 0 & 0 \\
-\lambda S_{1,z}^{\rm (ss)} & -\lambda S_{1,z}^{\rm (ss)} & \omega_a & 0 & -\lambda (a_{\rm ss}+a_{\rm ss}^*) & 0 & 0 & 0 \\
\lambda S_{1,y}^{\rm (ss)} & \lambda S_{1,y}^{\rm (ss)} & 0 & \lambda (a_{\rm ss}+a_{\rm ss}^*) & 0 & 0 & 0 & 0 \\
0 & 0 & 0 & 0 & 0 & 0 & -\omega_a & 0 \\
\lambda S_{2,z}^{\rm (ss)} & \lambda S_{2,z}^{\rm (ss)} & 0 & 0 & 0 & \omega_a & 0 & \lambda (a_{\rm ss}+a_{\rm ss}^*)  \\
-\lambda S_{2,y}^{\rm (ss)} & -\lambda S_{2,y}^{\rm (ss)} & 0 & 0 & 0 & 0 & -\lambda (a_{\rm ss}+a_{\rm ss}^*) & 0
\end{pmatrix}.
\end{align}
Recall that $S_{l,y}^{\rm (ss)}=0$ in all steady states.

%=========================================================================
\subsection{Semi-classical dynamics}

%--------Figure------------ 
\begin{figure}[t!]
\centering
\includegraphics [width=0.95\textwidth]{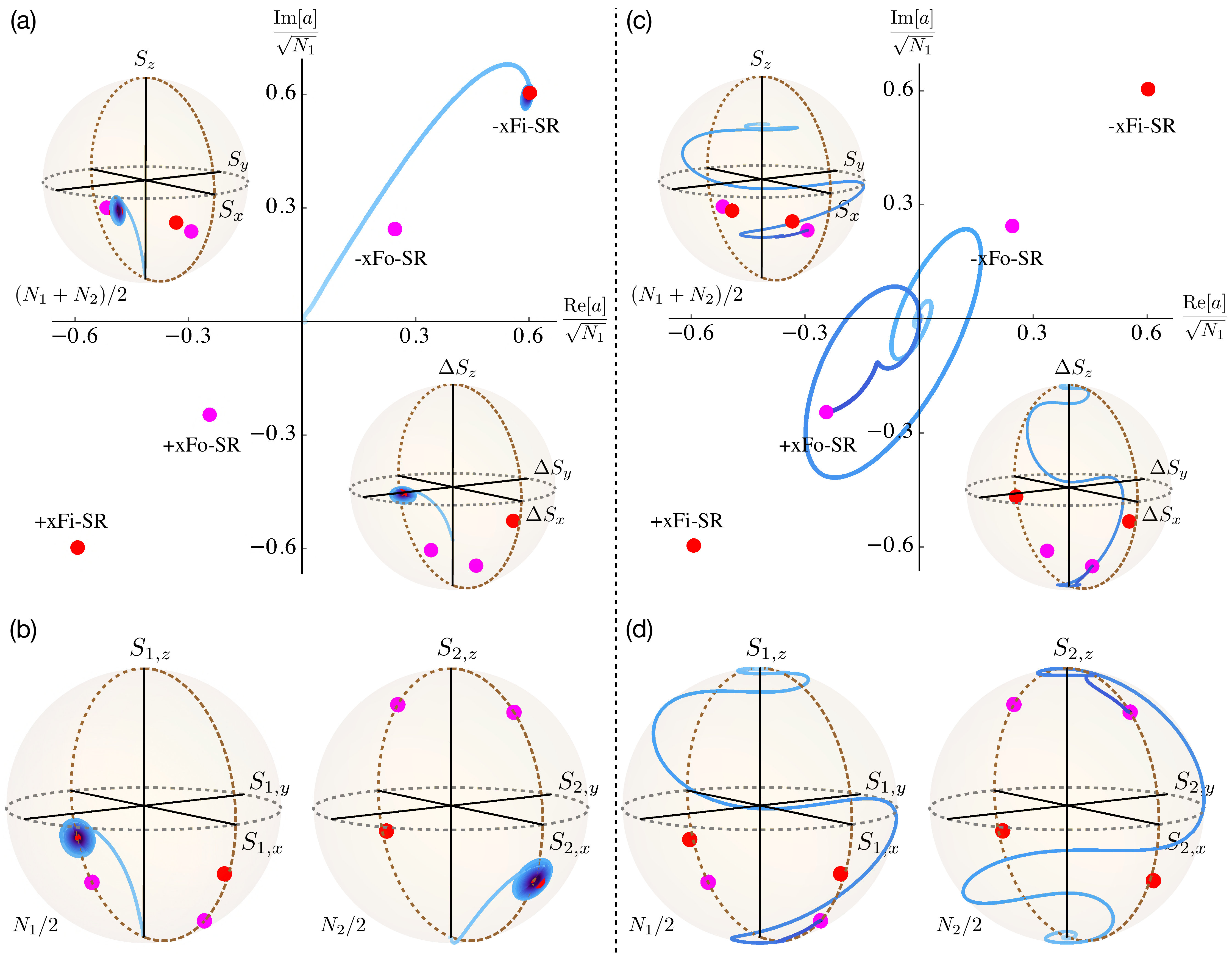}
\caption{The nonequilibrium dynamics of the system. 
The system is prepared in the initial state (a,b) $-$zFo-N, i.e., $\mathbf{S}_1=(0,0,-1)N_1/2$ and 
$\mathbf{S}_2=(0,0,-1)N_2/2$, and (c,d) +zFi-N,  i.e., $\mathbf{S}_1=(0,0,1)N_1/2$ and 
$\mathbf{S}_2=(0,0,-1)N_2/2$, with a small fluctuation seed for the cavity field $a=\delta a$.
The ensuing dynamics of the system are displayed in the phase space of the cavity-field
amplitude $a$ (a,c) and the $\mathbf{S}_1$ and $\mathbf{S}_2$ spin Bloch spheres (b,d).
The insets in panels (a,c) exhibit the dynamics in the
the Bloch spheres of the total $\mathbf{S}$ and staggered $\Delta\mathbf{S}$ spins. 
The color gradient from light blue to darker blue in trajectories indicates schematically the arrow of the time evolution.
The dynamic evolutions lead to the stable fixed point $-$xFi-SR in the first case (a,b), and to the +xFo-SR 
in the second case (c,d). However, in the first case trajectories are slowly inward-moving spirals and the time
scale to reach the fixed point is much longer than the second case. Therefore, $\pm$xFi-SR can be 
viewed as stable focuses, while $\pm$xFo-SR as stable nodes in this parameter regime.
The radii of the Bloch spheres are indicated in each case, where 
$\mathbf{S}$ and $\Delta\mathbf{S}$ have the same radius $(N_1+N_2)/2$.
Here, the parameters are the same as Fig.~\ref{fig:SC-dynamics-LC} in the main text.
} 
\label{fig_SM:SC-dynammics_xFo_xFi}
\end{figure}

Here in Fig.~\ref{fig_SM:SC-dynammics_xFo_xFi} we present the semi-classical dynamics of the system prepared in the initial state (a,b) $-$zFo-N, i.e., $\mathbf{S}_1=(0,0,-1)N_1/2$ and $\mathbf{S}_2=(0,0,-1)N_2/2$, and (c,d) +zFi-N,  i.e., $\mathbf{S}_1=(0,0,1)N_1/2$ and 
$\mathbf{S}_2=(0,0,-1)N_2/2$, with a small fluctuation seed for the cavity field $a=\delta a$. The system is attracted to the -xFi-SR fixed point in the former case, and to +xFo-SR in the latter case. These indicate that the fixed points $-$zFo-N and +zFi-N are unstable in this parameter regime (recall, however, that +zFi-N  is always unstable) and the attractor of the long-time dynamics depends crucially on the initial state due to the multistability and the coexistent fixed points. In particular, the quantity $\tilde{\mathbf S}^2=(S_{1,x}-S_{2,x})^2+(S_{1,y}-S_{2,y})^2+(S_{1,z}+S_{2,z})^2$ must be conserved during the dynamics. This is indeed the reason why the dynamics starting from the initial state $-$zFo-N (+zFi-N) leads to the attractor -xFi-SR (+xFo-SR), since they lie in the same symmetry sector with ${\tilde{\mathbf S}}^2=(N_1+N_2)^2/4$ [${\tilde{\mathbf S}}^2=(N_1-N_2)^2/4$].

%=========================================================================
\subsection{Long-time quantum dynamics}

\begin{figure}[t!]
\begin{center}
\includegraphics [width=0.85\textwidth]{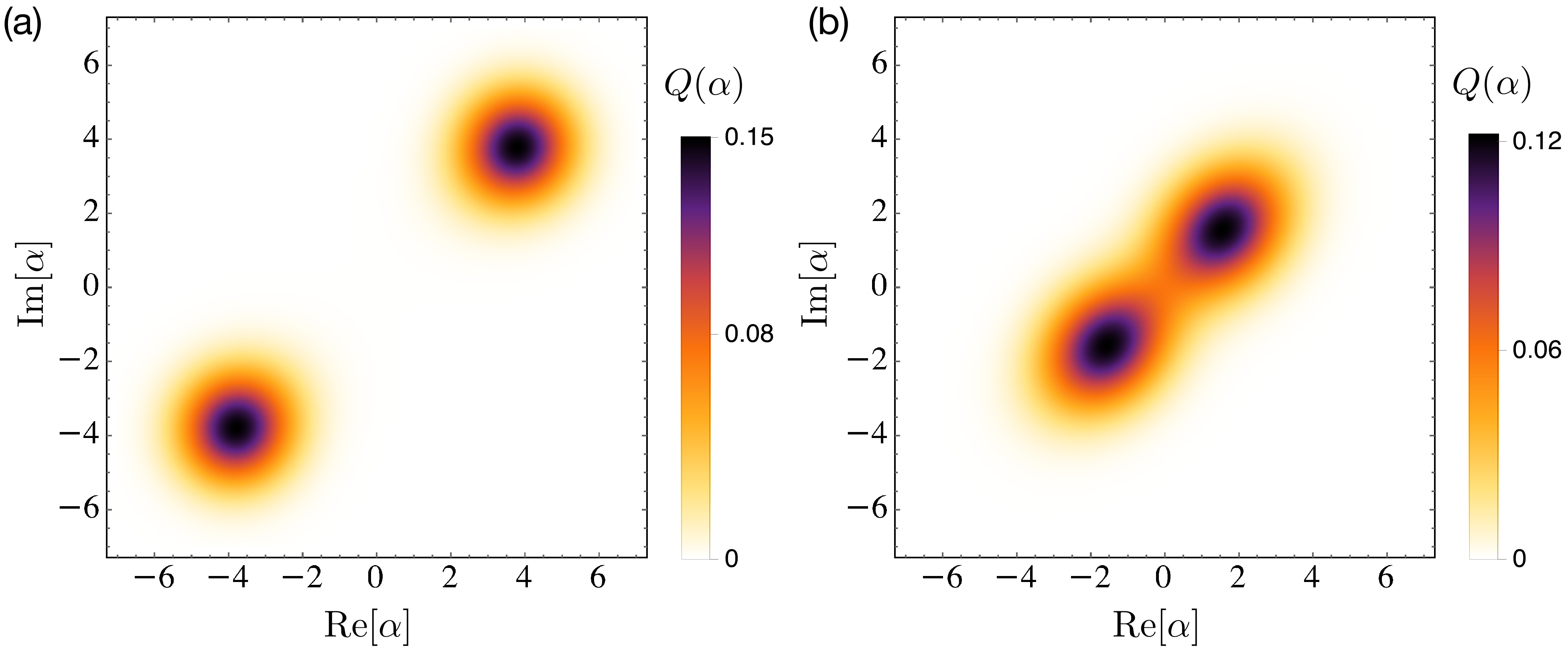}
\caption{The Husimi Q representation of the cavity field after long-time, quantum dynamics for different initial states: (a) $\ket{\downarrow}_1\otimes\ket{\downarrow}_2$ with $\langle\hat{\tilde{\mathbf S}}^2\rangle=56$, and (b) $\ket{\downarrow}_1\otimes\ket{\uparrow}_2$ with $\langle\hat{\tilde{\mathbf S}}^2\rangle=16$. The subscripts $\{1,2\}$ in the initial states refer to the first/second spin ensemble, respectively, and the initial coherent state is set to zero in both cases. The long-time Husimi Q representation corresponds to (a) $\pm$xFi-SR and (b) $\pm$xFo-SR. The parameters are the same as Fig.~\ref{fig:Q-function} in the main text.}
\label{fig_SM:q_function_Fo-Fi}
\end{center}
\end{figure}

Here we provide more examples of long-time quantum dynamics of the system obtained from the master equation. Figure~\ref{fig_SM:q_function_Fo-Fi} shows the Husimi $Q$ representation of the cavity field, defined as 
\begin{align}
Q(\alpha)=\frac{1}{\pi}\bra{\alpha}\hat{\rho}_{\rm cav}\ket{\alpha},
\end{align}
after long-time quantum dynamics for different initial states. Here, $\hat{\rho}_{\rm cav}=\text{Tr}_{\hat{\mathbf S}_1,\hat{\mathbf S}_2}(\hat{\rho})$ with $\text{Tr}_{\hat{\mathbf S}_1,\hat{\mathbf S}_2}$ being the partial trace over the two spin subsystems is the reduced density operator for the cavity subsystem and $\ket{\alpha}$ is the coherent state. The Husimi $Q$ representation corresponds to the $\pm$xFi-SR in Fig.~\ref{fig_SM:q_function_Fo-Fi}(a) and to the $\pm$xFo-SR in Fig.~\ref{fig_SM:q_function_Fo-Fi}(b).

%=========================================================================
\subsection{Possible experimental implementation}
\label{sec:SM-density-spin-self-ordering}

\begin{figure}[t!]
\centering
\includegraphics [width=0.95\textwidth]{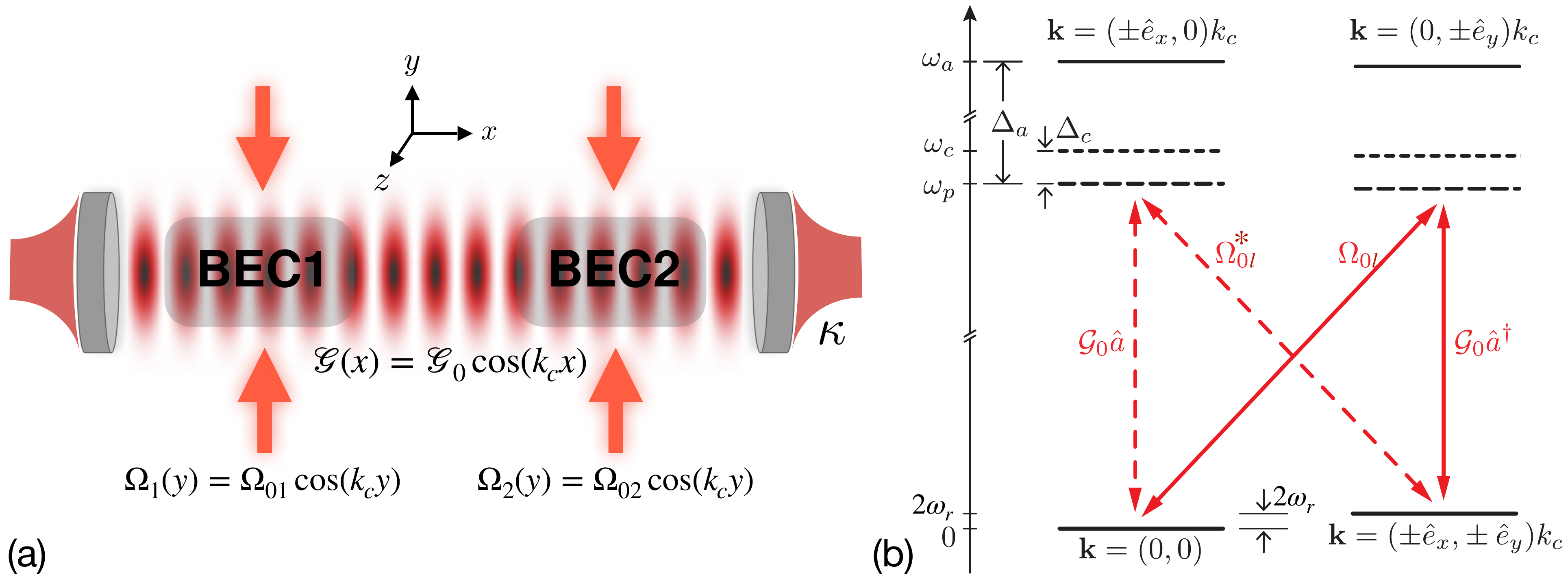}
\caption{Proposed implementation of the non-standard Dicke model.
(a) Two spatially separated BECs are coupled to the same standing-wave mode of a linear cavity. Furthermore, each BEC is driven in the transverse direction by a standing-wave pump laser. 
(b) Schematic representation of the atom-photon coupling. Both pump lasers and the cavity mode are all far detuned from any atomic transition. Hence, only low-lying atomic momentum states $(k_x,k_y)=(\pm k_c,\pm k_c)$ are populated out of the condensate $(k_x,k_y)=(0,0)$ during two-photon scattering processes between the pumps and the cavity mode.}
\label{fig_SM:cavity-BECs}
\end{figure}

Our proposed model can be implemented in state-of-the-art quantum-gas--cavity-QED setups. Consider two spatially separated, independent Bose-Einstein condensates (BECs) of the same atomic species inside a linear cavity~\cite{Joshi2015, Xu2019, Zhu2020}. The two BECs are driven independently by strengths $\Omega_{1,2}(y)=\Omega_{01,02}\cos(k_cy)$ in the transverse direction by pump lasers with the same angular frequency $\omega_p$ and coupled to a same cavity mode with strength $\mathcal{G}(x)=\mathcal{G}_0\cos(k_cx)$. The setup is depicted schematically in Fig.~\ref{fig_SM:cavity-BECs}(a). In the dispersive regime, due to the elastic photon scattering between the pumps and the cavity mode low-lying atomic momentum states $(k_x,k_y)=(\pm k_c,\pm k_c)$ in both BECs are populated out of the condensate $(k_x,k_y)=(0,0)$ as shown in Fig.~\ref{fig_SM:cavity-BECs}(b). One can define an independent spin-1/2 algebra using the low-lying momentum states of each BEC: $\hat{S}_{l,+}=\hat{S}_{l,-}^\dag=[\frac{1}{2}\sum_{n,m=\pm1}\hat{b}_{n,m}^{(l)\dag}]\hat{b}_{0,0}^{(l)}$ and $\hat{S}_{l,z}=[\sum_{n,m=\pm1}\hat{b}_{n,m}^{(l)\dag}\hat{b}_{n,m}^{(l)}-\hat{b}_{0,0}^{(l)\dag}\hat{b}_{0,0}^{(l)}]/2$ where $\hat{b}_{n,m}^{(l)}$ is the bosonic annihilation operator for the momentum state $(n,m)k_c$ in the $l$th BEC. It is then straightforward to map the system into this low-energy sector~\cite{Mivehvar2021},
\begin{align} \label{eq:SM-ns-Dicke-H-implementation}
\hat{H}_{\rm LE}=-\Delta_c\hat{a}^\dag\hat{a}+2\omega_r(\hat{S}_{1,z}+\hat{S}_{2,z})
+\frac{1}{2}(\hat{a}^\dag+\hat{a})(\eta_1\hat{S}_{1,x}+\eta_2\hat{S}_{2,x}),
\end{align} 
where $\Delta_c=\omega_p-\omega_c$ is the pump-cavity detuning, $\omega_r=k_c^2/2M$ the atomic recoil frequency, and $\eta_{l}=\Omega_{0l}\mathcal{G}_0/\Delta_a$ (with $\Delta_a=\omega_p-\omega_a$ being the pump-atom detuning and assuming $\{\Omega_{0l},\mathcal{G}_0\}\in\mathbb{R}$) the two-photon Raman-Rabi frequency of the $l$th BEC. Here, the optomechanical terms $\propto \mathcal{G}_0^2/\Delta_a$ have been assumed to be negligible and ignored. This is the general non-standard Dicke Hamiltonian given in Eq.~\eqref{eq:ns-Dicke-H} in the main text.

%--------Figure------------ 

\end{document}